\documentclass[12pt,twoside]{article}

\usepackage{amsmath,amssymb,amsthm,amscd,graphicx,color,accents,mathrsfs}

\usepackage[pagebackref=false,colorlinks=true,linkcolor=blue,citecolor=blue,urlcolor=blue]{hyperref}

\usepackage{url}

\numberwithin{equation}{section}

\theoremstyle{plain}

\DeclareMathOperator\E {\mathbb{E}}

\begin{document}

\title{\Large\textbf{A Continuous-Time Markov Chain Model for the Spread of COVID-19}}

\author{
{Armine Bagyan}\thanks{Department of Statistics, Pennsylvania State University, University Park, PA 16802, U.S.A. \ E-mail address: aub171@psu.edu}
\ {and Donald Richards}\thanks{Department of Statistics, Pennsylvania State University, University Park, PA 16802, U.S.A. \ E-mail address: richards@stat.psu.edu
\endgraf
\ $^\dag$Corresponding author.
}}

\date{\today}

\maketitle

\begin{abstract}
Since late 2019 the novel coronavirus, also known as COVID-19, has caused a pandemic that persists.  This paper shows how a continuous-time Markov chain model for the spread of COVID-19 can be used to explain, and justify to undergraduate students, strategies now being used in attempts to control the virus.  The material in the paper is written at the level of students who are taking an introductory course on the theory and applications of stochastic processes.  

\medskip
\noindent
{{\em Key words and phrases}. Cruise ship infections; Hypergeometric distribution; Infectious diseases; Kolmogorov forward equations; Omicron variant; Poisson process; Secondary attack rate; Stochastic modelling.}

\smallskip
\noindent
{{\em 2020 Mathematics Subject Classification}. Primary 60K50; Secondary 60G55.}

\smallskip
\noindent
{\em Running head}: Continuous-Time Markov Chain Modelling of COVID-19.

\end{abstract}

\section{Introduction}
\label{sec:intro}

Since late 2019, the novel coronavirus, also known as COVID-19, has caused a pandemic that shows little sign of abating.  National and international health organizations are now searching intensely for ways to control the spread of the virus in order to protect educational institutions, health organizations, and the broader societies and economies.  

In teaching introductory courses on the theory and applications of stochastic processes, we have shown students the wide applicability of these processes for modelling the spread of COVID-19; indeed we believe that the time remains opportune for teaching such applications.  In this paper, we provide an expanded classroom discussion on modelling the spread of the virus.  The exposition in the paper is at the level of an introductory course on the theory and applications of stochastic processes, and the material is accessible to students who have taken an introductory, calculus-based course in probability theory.

\section{A continuous-time Markov chain model for the spread of COVID-19}
\label{sec_CTMC_COVID}

Several textbooks on stochastic processes provide examples and homework problems which can be formulated as elementary models for the spread of infectious diseases.  For example, Ross \cite[p. 399, Exercise 6.5]{Ross} provides the following assignment: 

\bigskip
\bigskip

\begin{center}
\begin{tabular}{p{0.875\textwidth}}
There are $N$ individuals in a population, some of whom have a certain infection that spreads as follows.  Contacts between two members of this population occur in accordance with a Poisson process having rate $\lambda$.  When a contact occurs, it is equally likely to involve any of the $N$ pairs of individuals in the population.  If a contact involves an infected and a non-infected individual, then with probability $p$ the non-infected individual becomes infected.  Once infected, an individual remains infected throughout. Let $X(t)$ denote the number of infected members of the population at time $t$.
\end{tabular}
\end{center}

\bigskip
\bigskip

Students are asked to derive the expected waiting time, starting with a single infected individual, until all $N$ individuals have been infected.  In discussing this exercise with students we observe that the process $\{X(t), t \ge 0\}$ is a continuous-time Markov chain because contacts between individuals occur according to a Poisson process with constant rate, $\lambda$.  Also, each future state of the process $\{X(t), t \ge 0\}$ depends only on the current number of infected individuals and not on past numbers of infected individuals.  In addition, the process $\{X(t), t \ge 0\}$ is a \textit{pure-birth} process since infected individuals remain in the ``previously infected'' category.  

Suppose that at time $t$ there are $k$ infected individuals.  To calculate $\lambda_k$, the rate at which new infections now occur, suppose that two randomly chosen individuals come in contact.  Clearly, there are $\binom{k}{1}$ ways to choose one individual at random from the group of $k$ infected individuals; and there are $\binom{N-k}{1}$ ways to choose one individual at random from the group of $N-k$ non-infected individuals.  Also, there are $\binom{N}{2}$ ways to choose a pair of individuals at random from the population of $N$ individuals.  Therefore, the probability that the chosen pair consists of one infected and one non-infected individual is 
$$
\dfrac{\binom{k}{1} \cdot \binom{N-k}{1}}{\binom{N}{2}} = \dfrac{2k(N-k)}{N(N-1)}.
$$
This calculation can also be obtained by applying the hypergeometric probability distribution, which students will have studied in prior courses on probability theory.  
  
On multiplying the above result by $p$, we find that the probability that a contact between two randomly chosen individuals results in a newly infected individual is 
$$
p \cdot \dfrac{2k(N-k)}{N(N-1)} = \dfrac{2k(N-k)p}{N(N-1)}.
$$ 
Since contacts between individuals occur according to a Poisson process then 
\begin{equation}
\label{contact_rate}
\lambda_k = \lambda \cdot \dfrac{2k(N-k)p}{N(N-1)} = \dfrac{2k(N-k)\lambda p}{N(N-1)},
\end{equation}
where $k=1,\ldots,N-1$.  

Next, we recall for students the relationship between the exponential distribution and the Poisson process with rate $\lambda_k$: \textit{The waiting time for the next observation in a Poisson process has an exponential distribution with expected value} $\lambda_k^{-1}$ \cite[Subsection 5.3.3]{Ross}.  Therefore, the waiting time until all $N$ individuals are infected is the sum of the waiting times between successive infections.

Denote by $T$ the waiting time until all $N-1$ remaining uninfected individuals are infected. Then 
\begin{align}
\label{E_waiting_time}
\E(T) = \sum_{k=1}^{N-1} \lambda_k^{-1} &= \frac{(N-1)}{2p\lambda} \sum_{k=1}^{N-1} \dfrac{N}{k(N-k)} \nonumber \\
&= \frac{(N-1)}{2p\lambda} \sum_{k=1}^{N-1} \Big(\frac{1}{k} + \frac{1}{N-k}\Big) 
= \frac{(N-1)}{p\lambda} \sum_{k=1}^{N-1} \frac{1}{k},
\end{align}
where the last equality follows from the trivial identity, 
$$
\sum_{k=1}^{N-1} \frac{1}{k} = \sum_{k=1}^{N-1} \frac{1}{N-k}.
$$

\section{Application to the spread of COVID-19}
\label{sec_application}

Consider a ``population'' whose $N$ ``residents'' intermingle rapidly during the ongoing COVID-19 pandemic.  We model the residents' behavior using the scenario of the textbook problem stated in Section \ref{sec_CTMC_COVID}; the underlying assumptions are plausible if $N$, the size of the population, is large, say, in the thousands.  Examples of the scenario are $2,000$ frenzied fans attending a weekend-long, multi-team, tournament held in a high-school basketball arena; or 6,700 passengers who have set sail on a 10-day jaunt on a fully-packed cruise ship.  In particular, once such a fan or passenger becomes infected then the observed behavior of COVID-19 indicates that the person is likely to remain infected and infectious for the duration of the tournament or cruise, respectively.  

Denoting by $X(t)$ the number of infected individuals at time $t$, we model $X(t)$ as a continuous-time Markov chain process.  Then the expected waiting-time until all $N$ residents are infected is given by \eqref{E_waiting_time}.  

There are several variations on the above model for the spread of an infection.  We refer to Durrett \cite[p.~151]{Durrett} or Ross \cite[p.~360]{Ross} for a similar model, called the \textit{Yule process}, such that the rate of contact between two randomly chosen members of the population is $\lambda = N \mu$, where $\mu > 0$.  Then by \eqref{E_waiting_time}, it follows that 
\begin{equation}
\label{E_waiting_time_2}
\E(T) = \frac{(N-1)}{p\mu N} \sum_{k=1}^{N-1} \frac{1}{k}.
\end{equation}
For large values of $N$, we apply the well-known approximation for the harmonic series, 
$$
\sum_{k=1}^{N-1} \frac{1}{k} \simeq \ln N,
$$
and also apply $(N-1)/N \simeq 1$ to obtain the large-population approximation, 
\begin{equation}
\label{E_waiting_time_2_approx}
\E(T) \simeq \frac{\ln N}{p\mu}.
\end{equation}
In mathematical terminology, we have shown that $\E(T) = O(\ln N)$.  

The results \eqref{E_waiting_time_2} and \eqref{E_waiting_time_2_approx} enable students to insert values of $N$, $p$, and $\mu$ to compute the expected time when all members of a population will be infected.  For example, let $N=2,000$, the number of fans attending a weekend-long, multi-team, high-school basketball tournament.  We set $p=0.31$, a value which we choose due to data from Lyngse, \textit{et al.} \cite{Lyngse} on the COVID-19 Omicron variant's \textit{secondary attack rate}, i.e., the rate with which infections occur among contacts within the incubation period following exposure to a primary case.  Also, we set $\mu = 1$ per hour, meaning that each randomly chosen pair of fans is assumed to make contact, on average, once per hour.  Then we obtain $\E(T) \simeq 24.52$ hours.  

As another example, suppose that $N = 6,700$, the number of people on a fully-packed cruise ship now sailing on a $10$-day trip.  We again set $p = 0.31$, a value which is highly conservative for sea-going vessels \cite{Lee}.  Also, we set $\mu = 3$ per day, meaning that each randomly chosen pair of passengers is assumed to make contact, on average, thrice daily; this reflects the fact that passengers are likely to come in close contact at mealtimes.  Then we obtain $\E(T) \simeq 9.47$ days.  Bearing in mind that thousands of cruise ships have sailed during the pandemic, it is natural to expect that many ships will return with large numbers of COVID infections.  

We also observe from \eqref{E_waiting_time_2} and \eqref{E_waiting_time_2_approx} that $\E(T)$ is inversely proportional to $\mu$ and $p$.  Hence, the larger the value of $\mu$ or $p$, the smaller the expected waiting-time until all $N$ individuals are infected, and \textit{vice versa}.  This result will justify to students the importance of social-distancing as a tool for decreasing $\mu$; cf., Oraby, \textit{et al.} \cite{Oraby}.  

Further, students will also deduce from \eqref{E_waiting_time_2_approx} that vaccination and masking programs that lower the value of $p$, the probability that a new infection occurs, also will lead to larger values of $\E(T)$.  Therefore, \eqref{E_waiting_time_2} and \eqref{E_waiting_time_2_approx} underscore the importance of social-distancing and vaccination strategies in efforts to end the pandemic.

\section{Conclusions}
\label{sec_conclusions}

The model described in this paper is based on introductory methods in the theory of stochastic processes.  And yet it is worthy of note that the $O(\ln N)$  phenomenon appears in more complex stochastic models for the spread of epidemics; cf., Barbour \cite{Barbour}, Isham \cite{Isham}.  Although policy-makers will base their decisions on more sophisticated models for the spread of COVID-19, it will be thought-provoking for students to observe that their classroom models are able to deduce results consistent with deeper analyses.  

The analysis provided in Section \ref{sec_application} also enables students to begin their own research on the subject.  By considering the Kolmogorov forward equations, students can verify well-known formulas (Ross \cite[p.~367]{Ross}) for the transition probabilities for the process $\{X(t), t \ge 0\}$ and simulate numerically the corresponding sample paths (Allen \cite{Allen}).  These results can be used to demonstrate that even if $p$ and $\mu$ are small, the process can exhibit repeated outbreaks for very long periods of time (Isham \cite{Isham}).  

The preceding discussion pertains to the case in which $\lambda_k = ck$ where $c$ is a constant; see \eqref{contact_rate}.  Similar analyses will enable students to study the consequences for the population of other cases.  For example, suppose that individuals intermingle so rapidly that the rate of contact at state $k$ changes to $\lambda_k = c k^2$.  Students will discover that 
$$
\E(T) = \sum_{k=1}^N \lambda_k^{-1} = \frac{1}{c} \sum_{k=1}^N \frac{1}{k^2},
$$
which remains finite as $N \to \infty$.  By further probabilistic analysis, students can determine that $T$ itself remains finite, almost surely.  Computer simulations can then be used to supplement classroom discussions of the practical consequences of these results and to show visually the dramatic implications of $T$ being finite, almost surely.  As a practical matter, students will deduce that rapid intermingling can cause an entire country, \textit{regardless of the population size}, to be completely overrun with COVID-19 infections in a finite time.  

On the other hand, suppose that policies that reduce the rate of community spread were to result in new-infection rates of the form $\lambda_k = c/k^2$.  Then 
$$
\E(T) = \sum_{k=1}^N \lambda_k^{-1} = \frac{1}{c} \sum_{k=1}^N k^2 = O(N^3).
$$
This indicates that large populations especially are likely to benefit enormously from such spread-reduction measures, for the expected time-to-overrun will far exceed the population size.  

In conclusion, the model considered here is a good pedagogical starting point for students to apply their coursework to COVID-19 data.  Further, the present time is opportune for motivating students and faculty to read ongoing research on modelling the spread of COVID-19 and other infectious diseases; cf.,  Calleri, \textit{et al.} \cite{Calleri}, Oraby, \textit{et al.} \cite{Oraby}.

\bigskip
\bigskip

\noindent
\textbf{Disclosure statement}.  No financial support was received to underwrite the preparation of this manuscript, and the authors have no relevant financial or non-financial interests to disclose.

\bigskip

\noindent
\textbf{Data Availability Statement}: Data sharing is not applicable to this article as no datasets were generated or analysed during the current study.


\end{document}